\RequirePackage{lineno}
\documentclass[aps,prl,twocolumn,superscriptaddress,linenumbers]{revtex4}
\usepackage{ulem}
\usepackage{mathrsfs,amssymb,graphics,subfigure,threeparttable}
\usepackage{grap hicx}
\usepackage{amssymb}
\usepackage{enumerate}
\usepackage{amsmath}
\usepackage{fancyhdr}
\usepackage{bm}
\usepackage[squaren]{SIunits}
\usepackage{xcolor}
\usepackage{changes}
\usepackage{soul}
\usepackage{multirow}
\usepackage{hyperref}
\hypersetup{
    colorlinks=true,
    linkcolor=blue,
    filecolor=black,      
    citecolor=blue,
    urlcolor=blue,
    }

\begin{document}


\title{On the generation of ultra-bright and low energy spread electron beams in laser wakefield acceleration in a uniform plasma}



\author{Xinlu Xu}
\email[]{xuxinlu@pku.edu.cn}
\affiliation{State Key Laboratory of Nuclear Physics and Technology, and Key Laboratory of HEDP of the Ministry of Education, CAPT, Peking University, Beijing, 100871, China}
\affiliation{Beijing Laser Acceleration Innovation Center, Beijing 100871, China}
\author{Thamine N. Dalichaouch}
\affiliation{Department of Physics and Astronomy, University of California Los Angeles, Los Angeles, CA 90095, USA}
\author{Jiaxin Liu}
\affiliation{State Key Laboratory of Nuclear Physics and Technology, and Key Laboratory of HEDP of the Ministry of Education, CAPT, Peking University, Beijing, 100871, China}
\author{Qianyi Ma}
\affiliation{State Key Laboratory of Nuclear Physics and Technology, and Key Laboratory of HEDP of the Ministry of Education, CAPT, Peking University, Beijing, 100871, China}
\author{Jacob Pierce}
\affiliation{Department of Physics and Astronomy, University of California Los Angeles, Los Angeles, CA 90095, USA}
\author{Kyle Miller}
\affiliation{University of Rochester, Laboratory for Laser Energetics, Rochester, NY 14623, USA}
\author{Xueqing Yan}
\affiliation{State Key Laboratory of Nuclear Physics and Technology, and Key Laboratory of HEDP of the Ministry of Education, CAPT, Peking University, Beijing, 100871, China}
\affiliation{Beijing Laser Acceleration Innovation Center, Beijing 100871, China}
\affiliation{Collaborative Innovation Center of Extreme Optics, Shanxi University, Shanxi 030006, China}
\author{Warren B. Mori}
\affiliation{Department of Physics and Astronomy, University of California Los Angeles, Los Angeles, CA 90095, USA}
\affiliation{Department of Electrical Engineering, University of California, Los Angeles, California 90095, USA}

\date{\today}

\begin{abstract} 
The quality of electron beams produced from plasma-based accelerators, i.e., normalized brightness and energy spread, has made transformative progress in the past several decades in both simulation and experiment. Recently, full-scale particle-in-cell (PIC) simulations have shown that electron beams with unprecedented brightness ($10^{20}\sim10^{21}~\ampere/\meter^2/\rad^2$) and $0.1\sim 1~\mega\electronvolt$ energy spread can be produced through controlled injection in a slowly expanding bubble that arises when a particle beam or laser pulse propagates in density gradient, or when a particle beam self-focuses in uniform plasma or has a superluminal flying focus. However, in previous simulations of work on self-injection triggered by an evolving laser driver in a uniform plasma, the resulting beams did not exhibit comparable brightnesses and energy spreads. Here, we demonstrate through the use of large-scale high-fidelity PIC simulations that a slowly expanding bubble driven by a laser pulse in a uniform plasma can indeed produce self-injected electron beams with similar brightnesses and energy spreads as for an evolving bubble driven by an electron beam driver. We consider laser spot sizes roughly equal to the matched spot sizes in a uniform plasma and find that the evolution of the bubble occurs naturally through the evolution of the laser. The effects of the electron beam quality on the choice of physical as well as numerical parameters, e.g. grid sizes and field solvers used in the PIC simulations are presented. It is found that this original and simplest injection scheme can produce electron beams with beam quality exceeding that of the more recent concepts.  
\end{abstract}


\maketitle



\section{I. Introduction}
Critical metrics for accessing the capability of a particle accelerator are related to the quality of the beams they can deliver. Several macro or ensemble averaged quantities of importance to quantify the beams include the energy $E_\mathrm{b}$, energy spread $\sigma_\mathrm{E_\mathrm{b}}$, current $I$, normalized emittance $\epsilon_\mathrm{N}$, duration $\tau_\mathrm{b}$ and repetition rate $f$. As a promising novel acceleration method, plasma-based acceleration (PBA) driven by an intense laser pulse or particle beam can sustain ultrahigh acceleration gradients ($10\sim100~\giga\volt\per\meter$) within the acceleration medium - a fully ionized plasma \cite{PhysRevLett.43.267, chen1985acceleration, esarey2009physics, joshi2020perspectives}. Generation of high-quality electron beams from PBA is critical for its development as such beams could transform applications, such as X-ray free-electron lasers (XFELs) \cite{pellegrini2016physics, kim2017synchrotron} and TeV-class colliders \cite{benedetti2022linear}. Both of these applications have stringent requirements for the beam quality. Although high acceleration gradients were demonstrated experimentally in the early stages of PBA research \cite{clayton1993ultrahigh, everett1994trapped, modena1995electron}, the generated electron beams were characterized by large divergences and emittances, and Maxwellian energy distributions, i.e., $100\%$ energy spread \cite{modena1995electron, umstadter1996nonlinear, wagner1997electron, malka2002electron}. In 2004, three groups \cite{mangles2004monoenergetic, geddes2004high, faure2004laser} produced monoenergetic $\sim$100 MeV beams with a few percent energy spread and several mrad divergence by shooting a ultrashort ($30\sim 60~\femto\second$) laser pulse with $\sim$joule energy into a plasma with $\sim10^{19}~\centi\meter^{-3}$ density. However, the beam qualities had large shot-to-shot fluctuations due to the variation of the laser and plasma parameters when operating at the relatively high plasma densities. 

In order to improve the stability and reproducibility of the beams produced from PBA, a number of controllable injection schemes have been proposed during the last two decades. These schemes utilize a variety of physical mechanisms, such as additional lasers \cite{umstadter1996laser, esarey1997electron, hemker1998computer, fubiani2004beat, kotaki2004head, faure2006controlled, PhysRevLett.102.065001, lehe2013optical, PhysRevLett.128.164801}, an external magnetic field \cite{vieira2011magnetic}, a plasma density gradient \cite{PhysRevE.58.R5257, PhysRevLett.86.1011, PhysRevLett.100.215004, gonsalves2011tunable, PhysRevLett.110.185006, li2013dense}, or the vast difference in the ionization potentials of electrons between different shells of atoms \cite{PhysRevLett.82.1688, chen2006electron, PhysRevLett.98.084801, rowlands2008laser, PhysRevLett.104.025003, PhysRevLett.104.025004, PhysRevLett.105.105003, PhysRevLett.107.035001, PhysRevLett.107.045001, maier2020decoding}. 

To date, the representative mechanisms which have produced the best experimental and simulation results are ionization injection and density downramp injection in the nonlinear blowout regime. In ionization injection, electrons with high ionization potentials are released inside the wake and these electrons are more easily trapped (injected) than background electrons. In density downramp injection, a negative plasma density gradient is used to reduce the phase velocity of the wake by gradually increasing the wavelength of the wake and trigger injection of energetic plasma sheath electrons. The phase space dynamics of the injected electrons in these two schemes have been thoroughly investigated \cite{PhysRevLett.112.035003, xu2017high} and many variations have been proposed to further improve the generated beam quality \cite{xu2017high, PhysRevLett.108.035001, PhysRevLett.111.015003, PhysRevLett.111.155004, PhysRevLett.111.245003, PhysRevLett.112.125001, PhysRevSTAB.17.061301, zeng2015multichromatic, xu2016nanoscale, tooley2017towards, deng2019generation,  PhysRevAccelBeams.23.021304, PhysRevAccelBeams.25.011302, li2022ultra, xu2022generation, wang2022MRE}. 

Recent full-scale high-fidelity particle-in-cell (PIC) simulations have shown that downramp injection and other schemes that gradually increase the wavelength in the nonlinear blowout regime \cite{xu2017high, PhysRevAccelBeams.23.021304, li2022ultra} can produce beams with unprecedented brightness $B = \frac{2I}{\epsilon_\mathrm{N}^2} = 10^{20}\sim10^{21}~\ampere/\meter^2/\rad^2$ ($I\sim 20 ~\kilo\ampere$ and $\epsilon_\mathrm{N}\sim 10 ~\nano\meter$) and low slice energy spreads of $0.1\sim1~\mega\electronvolt$. The large energy chirp formed during the injection process can be compensated by the chirp of the acceleration gradient of the nonlinear wake \cite{PhysRevLett.96.165002, PhysRevLett.101.145002} during the subsequent acceleration process \cite{kalmykov2009electron, kalmykov2011electron, zhang2016energy, xu2017high, manahan2017single, kirchen2021optimal}. As a result, the beams can achieve $\sim0.1\%$ or even smaller relative projected energy spreads after they are boosted to GeV-class or higher energies \cite{manahan2017single, pousa2019compact, PhysRevAccelBeams.23.021304, ke2021near}.  

Beside these controllable injection schemes, self-injection induced by the evolution of an intense laser driver \cite{pukhov2002laser, tsung2004near, xu2005electron, PhysRevSTAB.10.061301, kalmykov2009electron, kostyukov2009electron, kalmykov2011electron, martins2010exploring} in a uniform plasma with density around $10^{17}\sim 10^{18}~\centi\meter^{-3}$ has been commonly used in experiments to generate GeV-class electrons due to its simplicity \cite{leemans2006gev, PhysRevLett.103.035002, PhysRevLett.103.215006, hafz2008stable, wang2013quasi, PhysRevLett.111.165002, leemans2014multi, PhysRevLett.122.084801, miao2022multi, aniculaesei2022high}. Notably, this self-injection mechanism was used to generate beams with energies above the GeV barrier \cite{leemans2006gev} and, more recently, up to 8 GeV \cite{PhysRevLett.122.084801}. However, these beams were characterized by relatively poor quality, i.e., large energy spreads ($\frac{ \sigma_{\mathrm{E_b}} }{ E_\mathrm{b}} \gg 1\%$) and large emittance ($\sim \micro\meter$), in both PIC simulations and experiments \cite{tsung2004near, leemans2006gev, PhysRevLett.103.035002, PhysRevLett.103.215006, hafz2008stable, kalmykov2009electron, kostyukov2009electron, martins2010exploring, kalmykov2011electron, wang2013quasi, PhysRevLett.111.165002, leemans2014multi, PhysRevLett.122.084801, corde2013observation, davidson2018optimizing}. 

In this paper, we demonstrate using large-scale high-fidelity PIC simulations and theoretical analysis that the dynamics of self-injection induced by the evolution of a short pulse laser are fundamentally the same as those observed in downramp injection. Thus, the self-injected electrons can achieve similar beam quality as from downramp injection, i.e., $10^{20}\sim10^{21}~\ampere/\meter^2/\rad^2$ brightness, sub-MeV slice energy spread, and $\lesssim 0.1\%$ relative projected energy spread after the beam is boosted to GeV-class energies. In Sec. II, we compare simulation results for both a uniform plasma and a plasma with a density downramp to show the similarity between these two regimes. A GeV-class beam with $0.3\times10^{20}~\ampere/\meter^2/\rad^2$ and $0.2\%$ relative projected energy spread is produced in the example. In Sec. III, we explore the potential of self-injection in a uniform plasma. We find that for diffraction-limited laser beams and perfectly uniform plasmas this simple scheme can generate beams with ultrahigh brightness ($\sim 10^{21}~\ampere/\meter^2/\rad^2$) and ultrahigh current ($\sim 100~\kilo\ampere$) by operating at $\sim10^{19}~\centi\meter^{-3}$ plasma densities. Possible physical factors which prevent the generation of such high-quality beams in experiments are briefly discussed in Sec. IV and a summary of our findings is given in Sec. V. The effects of numerical resolution and the choice of the electromagnetic field solver on the injected beams' quality are studied in the Appendix A.

\section{II. Common dynamics of self injection in uniform plasma and density downramps}
We model the self-injection of plasma electrons in a laser wakefield accelerator using the quasi-three-dimensional (Q3D) version \cite{davidson2015implementation} of the PIC code OSIRIS \cite{fonseca2002high} with recently developed high-fidelity Maxwell solvers \cite{li2017controlling, xu2020numerical}. A plasma column with density $n_\mathrm{p0}=2\times 10^{18}~\centi\meter^{-3}$ and isotropic electron temperature $T=[0.1~\electronvolt, 0.1~\electronvolt, 0.1~\electronvolt]$ is initialized at the beginning of the simulation. An 800 nm, 4.3 J laser pulse with a peak power $P=155~\tera\watt$ and a full-width half maximum (FWHM) pulse duration $\tau_\mathrm{FWHM}=38.7~\femto\second ~(3.1~\omega_\mathrm{p0}^{-1})$ is incident on the fully ionized plasma, where $\omega_\mathrm{p0}=\sqrt{\frac{n_\mathrm{p0}e^2}{m\epsilon_0}}$ is the plasma frequency, $m$ and $e$ are the electron mass and charge, and $\epsilon_0$ is the vacuum permittivity. The laser has a diffraction-limited Gaussian transverse profile and is focused at the start of the plasma ($z=0$) with a spot size $w_0=16.9~\micro\meter~(4.5 k_\mathrm{p0}^{-1})$ (the field profile at focus is $e^{-r^2/w_0^2}$), where $k_\mathrm{p0}=\frac{\omega_\mathrm{p0}}{c}$ is the plasma wavenumber and $c$ is the speed of light in vacuum. The normalized vector potential of the laser is $a_0\equiv\frac{eA_\mathrm{L}}{mc^2}=\frac{eE_\mathrm{L}}{mc\omega_0}=4$ at its focus, where $\omega_0$ is the laser frequency, $A_\mathrm{L}$ is the peak vector potential, and $E_\mathrm{L}$ is its peak electric field. The spot size used in the simulations is slightly larger than the matched spot size ($k_\mathrm{p0}w_\mathrm{0,match}=k_\mathrm{p0}r_\mathrm{b}=2\sqrt{a_0}=4$) suggested by the nonlinear plasma wave wake theory \cite{PhysRevLett.96.165002, PhysRevSTAB.10.061301}, where $r_\mathrm{b}$ is the radius of the blowout wake. The laser is linearly polarized along the $x$-direction. The grid sizes are chosen as $\mathrm{d}z=\mathrm{d}r=\frac{1}{512}k_\mathrm{p0}^{-1}$ to resolve the subtle physics involved in the injection process and the acceleration of the strongly focused injected electrons. Details of the simulation parameters can be found in Appendix B.

\begin{figure}[bp]
\includegraphics[width=0.5 \textwidth]{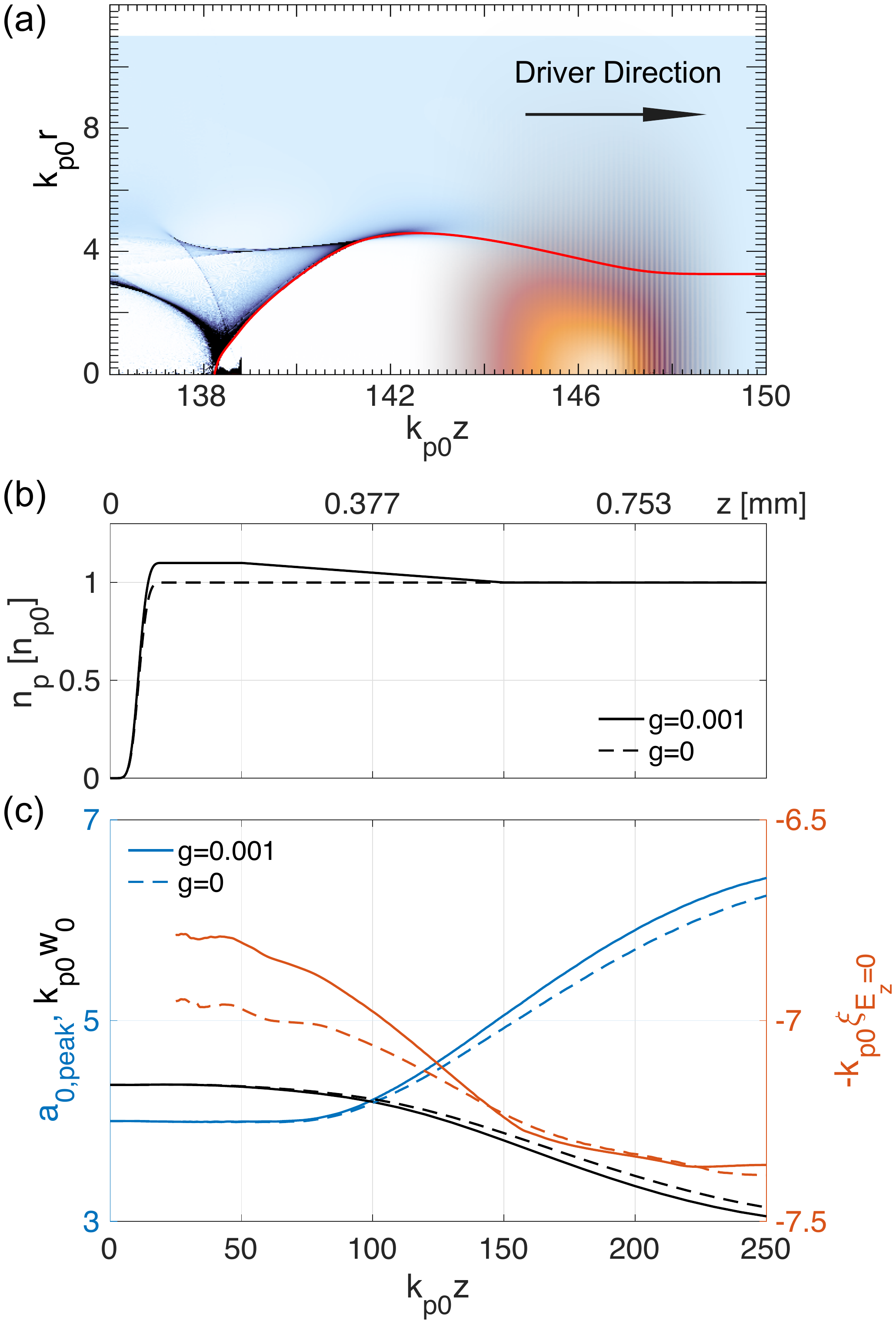}
\caption{\label{fig: concept} (a) The charge density distribution of a nonlinear plasma wake (white-blue-black) driven by a laser pulse (orange) at $\omega_\mathrm{p0}t=150$ in a uniform plasma. The trajectory of an injected electron (red line) is superimposed. (b) The plasma density profile. The downramp starts from $k_\mathrm{p0}z=50$ with density 1.1 $n_\mathrm{p0}$ and ends at $k_\mathrm{p0}z=150$ with density $n_\mathrm{p0}$. A super-gaussian upramp profile with order 4 is used between $z=0$ and $20~k_\mathrm{p0}^{-1}$. (c) The evolution of the peak $a_0$ (blue lines) and the projected spot size (black lines) of the laser pulse, and the position of the plasma wake where $E_\mathrm{z}=0$ (red lines).}
\end{figure}

As shown in Fig \ref{fig: concept}, we consider the self-injection for two cases: a linear plasma density downramp with normalized density gradient $g\equiv \frac{\Delta n / n_\mathrm{p0}}{ k_\mathrm{p0} L_\mathrm{ramp}}=0.001$ and a uniform plasma with $g=0$, where $\Delta n$ is the density drop across the ramp and $L_\mathrm{ramp}$ is the ramp length. When an intense laser pulse propagates inside a plasma (from left to right), the electrons are pushed outwards and forward from the pulse center. The ions then pull these electrons back causing them to form a narrow sheath that surrounds an ion column [Fig. \ref{fig: concept}(a)]. In general, when the laser's peak power exceeds the self-focusing critical power \cite{esarey1997self}, the plasma's refractive index is self-consistently modified to focus and guide the laser \cite{mori1997physics}. In the simulations conducted here, the peak power of the laser (155 $\tera\watt$) is much higher than the critical power for self-focusing in the plasma $P_\mathrm{c}\approx 17\left( \frac{\omega_0}{\omega_\mathrm{p0}} \right)^2 ~\giga\watt = 14.8~\tera\watt$ for the density in the plateau. Thus, the laser is self-focused and its peak field (projected spot size) increases (decreases) as shown in Fig \ref{fig: concept}(c). Despite the differences in the plasma density profiles, the evolution of the lasers are similar in these two cases. 
 
\begin{figure*}
\includegraphics[width=0.9\textwidth]{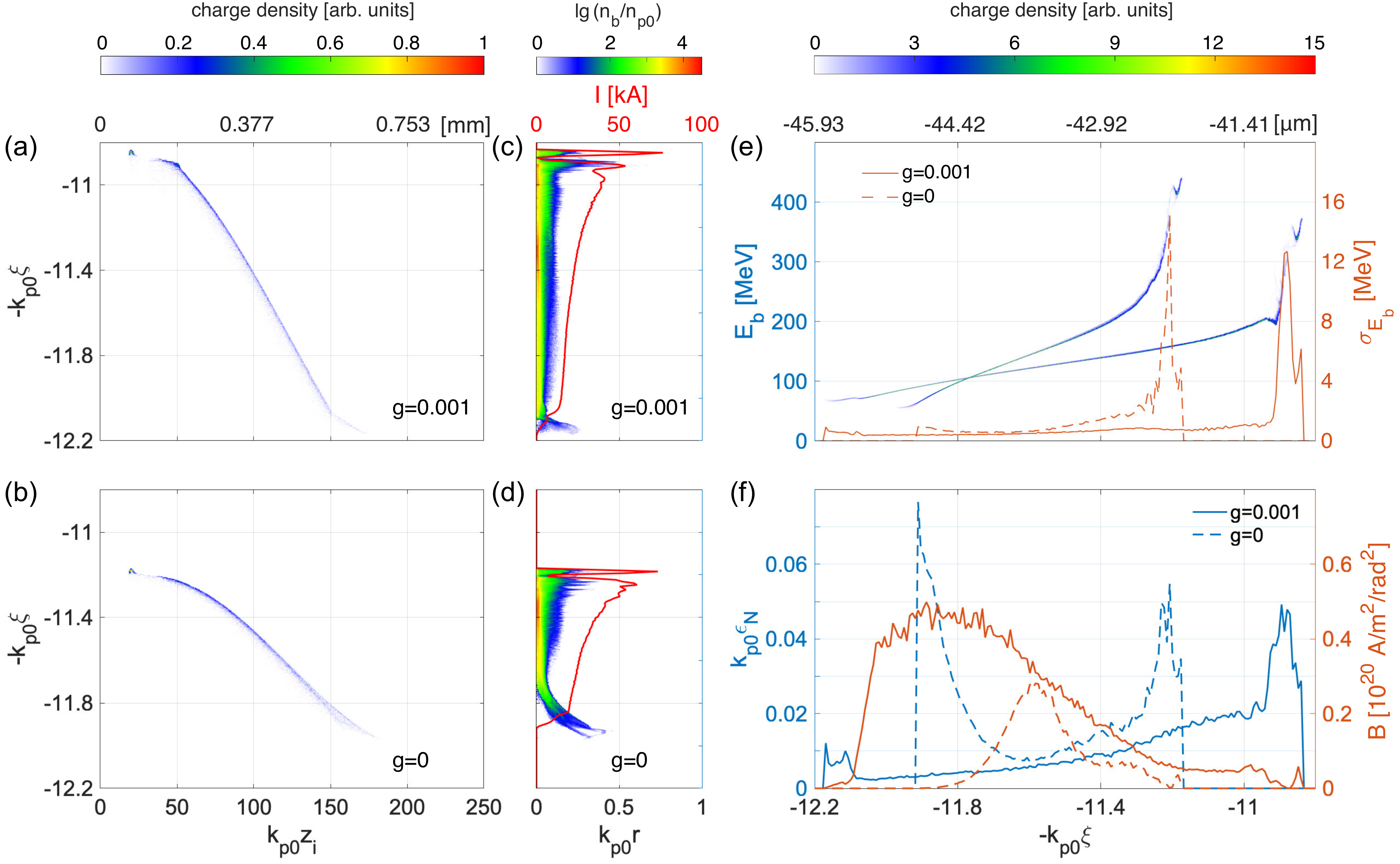}
\caption{\label{fig: downramp and no ramp} The initial $z_\mathrm{i}$ and the axial position after injection $\xi$ of the injected electrons for $g=0.001$ (a) and $g=0$ (b). The density distribution and the current (red lines) of the injected beam at $\omega_\mathrm{p0}t=250$ for $g=0.001$ (c) and $g=0$ (d). (e) The longitudinal phase space and the slice energy spread for both cases at $\omega_\mathrm{p0}t=250$. (f) The slice emittance along $x$-direction (blue lines) and the brightness (red lines) at $\omega_\mathrm{p0}t=250$. The duration of each slice is $\frac{1}{128}k_\mathrm{p0}^{-1}$ when calculating the slice properties.}
\end{figure*}

The wavelength of a laser-driven 3D nonlinear plasma wake depends on the spot size and the peak intensity of the laser driver in a complicated way, particularly when the beam is not well matched. In general, the wavelength tends to increase with both spot size and laser intensity. However, for the parameters studied here, the projected spot size decreases while the intensity increases through self-focusing in the ion channel. In this case, the increase in the intensity dominates, leading to a slow expansion of the wake size (wavelength). Previous simulations have also shown the wake expands as the laser diffracts for an initially tightly focused beam \cite{xu2005electron}. Since it is difficult to locate the precise end of the wake which is occupied by the self-injected electrons, we measure the axial position in the first wave bucket where $E_\mathrm{z}=0$ to quantify the expansion of the wake [red lines in Fig \ref{fig: concept}(c)]. Note the position where $E_\mathrm{z}=0$ is roughly the center of the wake and the rear of the wake expands with approximately twice this velocity. In the uniform plasma case, the self-focusing of the laser driver causes an expansion of the wake with a velocity, $v_{\phi,E_z=0}$, of $\sim 0.003c$ between $\sim50k_\mathrm{p0}^{-1}$ and $\sim 150k_\mathrm{p0}^{-1}$. The wake expands with a faster velocity of $\sim 0.005 c$ in the case with a density downramp which indicates the downramp speeds up the wake expansion.  

In a nonlinear wake, plasma electrons originating from $r_\mathrm{i} \sim \kappa r_\mathrm{b}$ form the high-density sheath surrounding the ion channel \cite{xu2017high} and gain large forward velocities when they move to the back of the wake, where $r_\mathrm{i}$ is the initial radial position of the electron and $\kappa$ is a value between $0.5$ and 1 which depends on the driver. If their forward velocities are faster than the phase velocity of the end of the wake, these sheath electrons would remain just inside at the end of the wake where they would then be accelerated continuously. The trajectory of a sample injected electron is superimposed on the wake shape as shown in Fig. \ref{fig: concept}(a).

Since the electrons are always injected at the rear of the continuously expanding wake, there is a mapping between the initial positions ($z_\mathrm{i}$) of the electrons and their axial positions inside the wake after injection ($\xi$) \cite{kalmykov2009electron, kostyukov2009electron, kalmykov2011electron, xu2017high} which can be seen in Figs. \ref{fig: downramp and no ramp}(a) and (b). The duration of the beam is roughly equal to the difference of the wake wavelength at the start and the end of the injection. The injected beam from a plasma downramp has a longer duration since the wake is smaller at the start of the injection where the plasma density is $1.1n_\mathrm{p0}$. The compression factor thus scales as $\sim \gamma_{\phi,E_z=0} ^{2}$ \cite{xu2017high, tooley2017towards}, where $\gamma_{\phi,E_z=0}=1/\sqrt{ 1 - v_{\phi, E_z=0}^2 /c^2 }$. Thus, there is a significant compression of the beams' duration during the injection, i.e., for these simulations the electrons initially distributed with a length of $\sim 100~k_\mathrm{p0}^{-1}$ and are compressed and form a beam with a duration of $\sim k_\mathrm{p0}^{-1}$. This enables the generation of beam currents of 10s of kA. As shown in Figs. \ref{fig: concept}(c) and (d), the current of the core of the beams are $\sim 30~\kilo\ampere$. Here the core of the beams is defined as part with brightness $\geq 10^{19}~\ampere/\meter^2/\rad^2$, which is $11.3<k_\mathrm{p0}\xi < 12.1$ for the $g=0.001$ case and $11.47<k_\mathrm{p0}\xi < 11.7$ for the $g=0$ case. The charge of the beam core is 180 $\pico\coulomb$ for $g=0.001$ and 76 $\pico\coulomb$ for $g=0$. We note that in addition to the continuous injection during the expansion of the wake, there is an isolated injection near the end of the upramp ($z=20 k_\mathrm{p0}^{-1}$) \cite{li2013dense}. This injection forms an attosecond peak with $\sim 70~\kilo\ampere$ current at the head of the beams in both cases \cite{li2013dense}. 

Based on analyzing numerous PIC simulations of downramp injection and its variants by the authors and others \cite{grebenyuk2014beam, xu2017high, PhysRevAccelBeams.23.021304, li2022ultra}, we have inferred that the current of the core of the beam scales as $I \sim \frac{\Lambda}{2} I_\mathrm{A} = I_\mathrm{d}$ for a beam-driven nonlinear wake and $ I \sim \frac{a_0}{2}I_\mathrm{A}$ for a laser-driven case (with a nearly matched spot size), where $\Lambda \equiv \frac{2I_\mathrm{d}}{I_\mathrm{A}}$, $I_\mathrm{d}$ is the peak current of the beam driver and $I_\mathrm{A} \approx 17~\kilo\ampere$ is the non-relativistic Alfven current. While we currently do not have simple arguments for obtaining these scaling laws, they may arise due to a fine balance between the initial injection and a subsequent quenching via self-beam loading, i.e., the injected electrons can modify the shape of the wake which reduces the forward velocity of the sheath electrons and hinders injection \cite{Thamine2022multisheath}. This observed scaling suggests that beams with hundreds of kA current and nanocoulombs of charge can be injected into a wake driven by an intense laser driver with $a_0 \gtrsim 10$. 

\begin{figure}[bp]
\includegraphics[width=0.5 \textwidth]{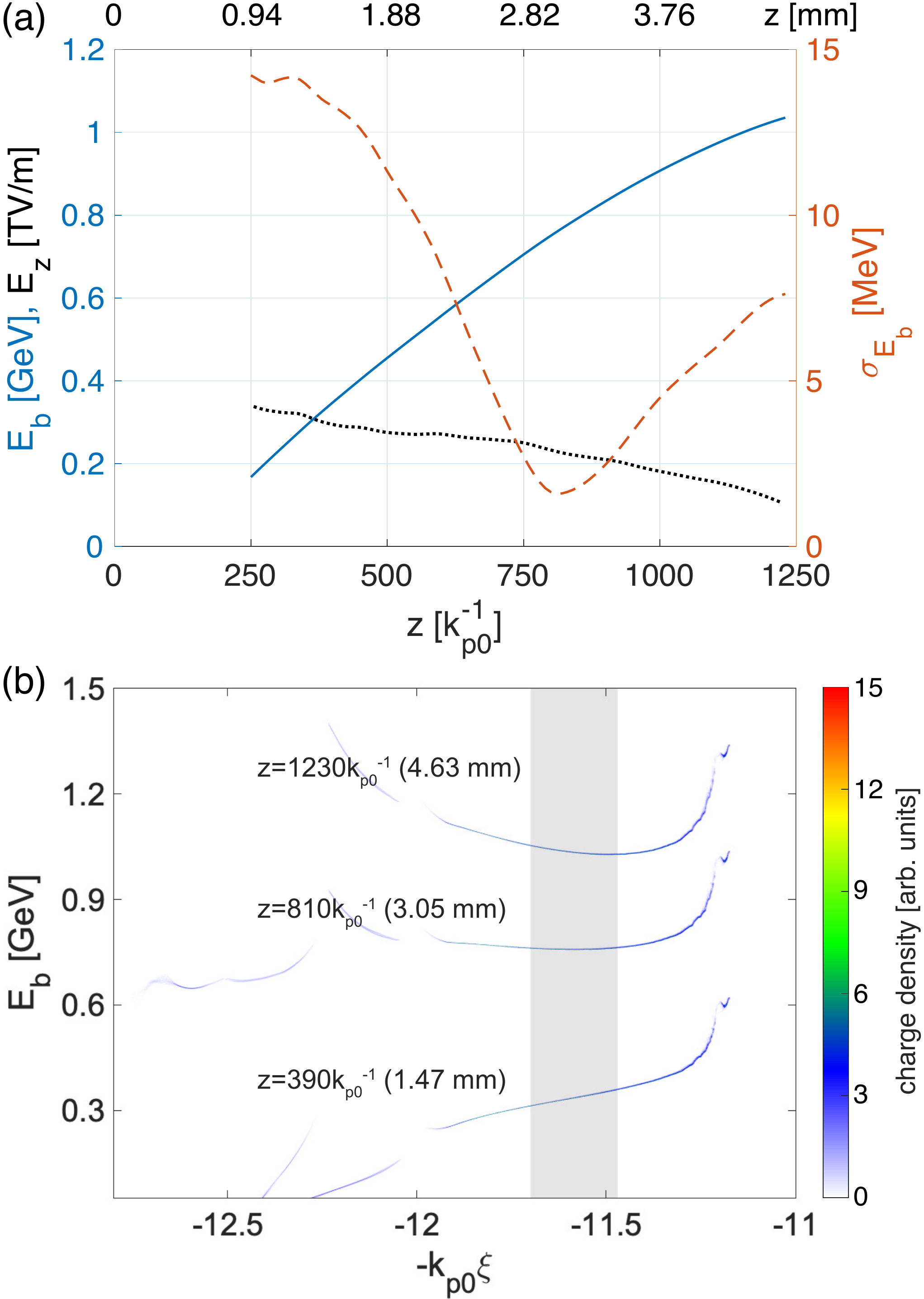}
\caption{\label{fig: longi_ps} (a) The evolution of the average energy (blue solid line) and the projected energy spread (red dashed line) of the core of the beam ($g=0$) and the acceleration gradient (black dotted line). (b) Longitudinal phase space of the injected electrons at three acceleration distances. The shadowed region represents the beam core. }
\end{figure}

The mapping between the initial positions of the electrons and their axial positions after injection leads to two consequences of the beams' energy distribution. The first is a low slice energy spread since the electrons in one axial slice originate from similar longitudinal locations and experience the same acceleration gradient after injection. Fig. \ref{fig: downramp and no ramp}(e) shows that the beams have a slice energy spread of $\sim 0.5~\mega\electronvolt$ except for near their heads. The second is a roughly linear energy chirp along the beam since the electrons at the beam head are injected earlier and accelerated over a longer distance. As shown in Fig. \ref{fig: downramp and no ramp}(e), the chirps at $\omega_\mathrm{p0}t=250$ are $174~\mega\electronvolt / (k_\mathrm{p0}\xi)~(46~\mega\electronvolt/\micro\meter)$ for $g=0.001$ and $423~\mega\electronvolt / (k_\mathrm{p0}\xi)~(112~\mega\electronvolt/\micro\meter)$ for $g=0$. 

These positive chirps can be compensated by the chirp of the acceleration gradient during the subsequent acceleration. The beam loaded wake has a lower gradient at the head and higher gradient at the rear. Thus there is an optimized acceleration distance where the beam can achieve a projected energy spread on the order of MeV \cite{xu2017high, ke2021near}. Due to the constant evolution of the laser driver, the acceleration gradient and its chirp change and we rely on simulations to find this optimized distance. We present the evolution of the average energy, the projected energy spread of the beam core ($11.47<k_\mathrm{p0}\xi < 11.7$) for $g=0$ and the acceleration gradient in Fig. \ref{fig: longi_ps}(a). The optimized acceleration distance occurs at $z=810k_\mathrm{p0}^{-1}~(3.05~\milli\meter)$ where the core of the beam achieves a projected energy spread (red dashed line) as low as 1.6 MeV while its average energy (blue solid line) is 0.76 GeV. The energy of the beam can still be boosted since there is $\sim75\%$ energy left in the laser pulse. We continue the simulation to $z=1230k_\mathrm{p0}^{-1}~(4.63~\milli\meter)$ where the core of the beam has 1.04 GeV energy and the laser pulse contains $\sim60\%$ of its initial energy. The average acceleration gradient drops from $\sim 0.35~\tera\volt\per\meter$ at $z=250k_\mathrm{p0}^{-1}~(0.94~\milli\meter)$ to $\sim0.1~\tera\volt\per\meter$ at $z=1230k_\mathrm{p0}^{-1}~(4.63~\milli\meter)$. A simulation with lower resolution ($\mathrm{d}z=\mathrm{d}r=\frac{1}{128}k_\mathrm{p0}^{-1}$) shows the beam energy reaches its maximum 1.40 GeV at $z=2440k_\mathrm{p0}^{-1}~(9.19~\milli\meter)$ and then starts to lose energy. The longitudinal phase space of the injected beam at three acceleration distances are shown in Fig. \ref{fig: longi_ps}(b) where the evolution of the energy chirp can be seen. Secondary and tertiary phases of injection occur around $z=300k_\mathrm{p0}^{-1}~(1.13~\milli\meter)$ and $750k_\mathrm{p0}^{-1}~(2.82~\milli\meter)$ which are characterized by low current, large emittance and large slice energy spreads. The aforementioned analysis and simulation results show that the injection mechanism and the longitudinal mapping are similar in a downramp and a uniform plasma. 

\begin{figure}[bp]
\includegraphics[width=0.5\textwidth]{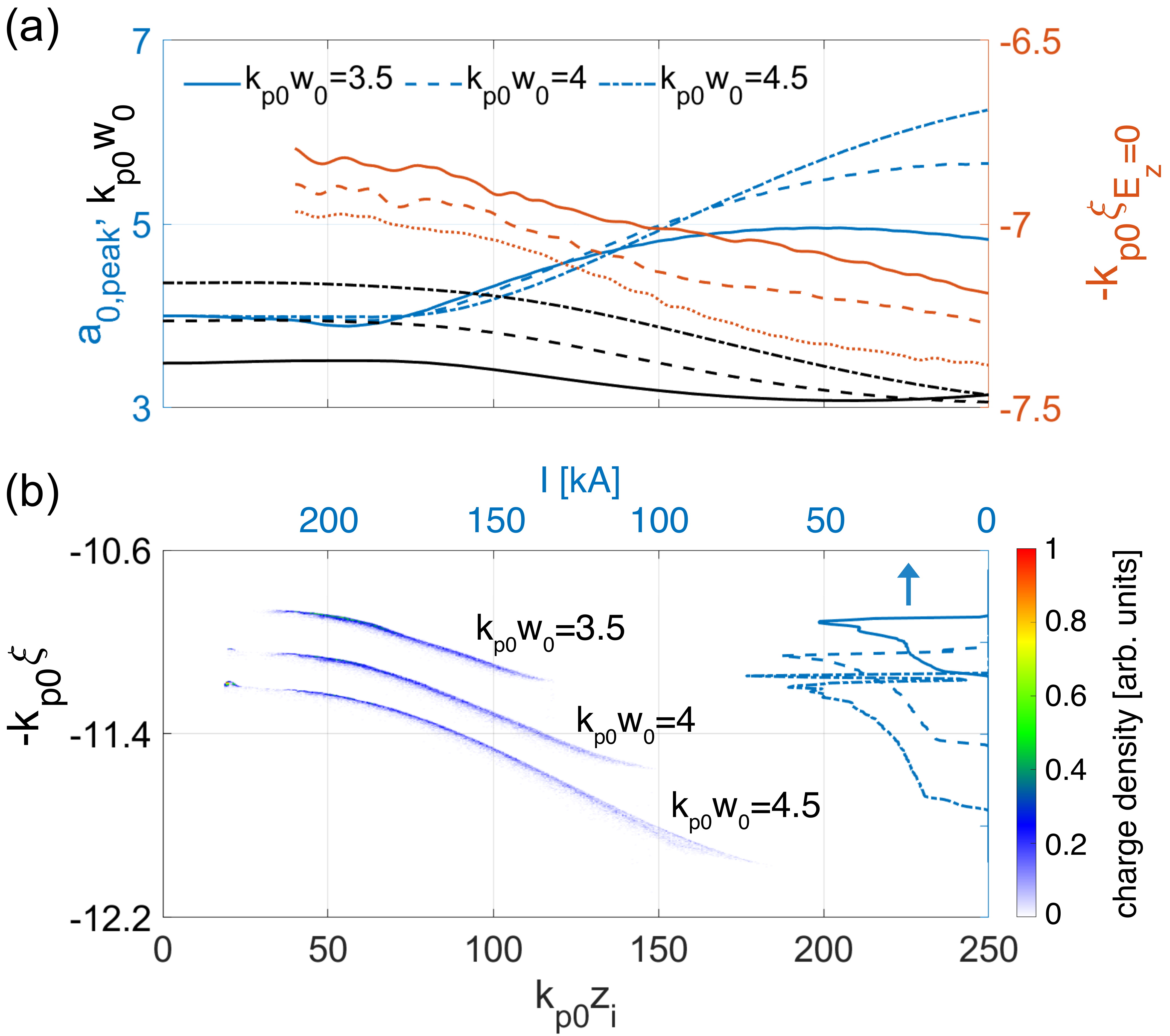}
\caption{\label{fig: w0 scan} (a) The evolution of the laser pulse driver and the plasma wake wave under different laser spot sizes: $k_\mathrm{p0}w_0=3.5, 4$ and 4.5. (b) The charge density distribution of the injected electrons and their current profiles for these cases.}
\end{figure}

As shown in Ref. \cite{xu2017high}, sheath electrons experience a transverse defocusing force from the nonlinear wake at the very rear of the channel prior to injection \cite{Thamine2022multisheath}. This transverse defocusing force reduces the transverse momentum of these electrons as they approach the axis leading to a beam with ultra-low emittance. The same dynamics occur for electrons injected from an elongating wake in a uniform plasma. The emittance of the beams along the laser polarization direction ($x$) are shown in Fig. 2(f) where the middle of the beams can achieve an emittance as low as $< 0.01~k_\mathrm{p0}^{-1}~(38~\nano\meter)$ while the head and the tail are characterized by a larger emittance of $0.01\sim 0.04 ~k_\mathrm{p0}^{-1}~(\sim 100~ \nano\meter)$. The relatively large emittance at the beam tail is due to the lack of symmetry of the injected electrons at the end of the injection process, i.e., the initial angular distribution of the injected electrons is asymmetric as shown in the Appendix C. The emittance along the other transverse direction has a similar profile. The peak brightness of the beams is $0.5\times 10^{20}~\ampere/\meter^2/\rad^2$ for $g=0.001$ and $0.3\times 10^{20}~\ampere/\meter^2/\rad^2$ for $g=0$. 

In principle, the injection of ultrahigh quality electrons in uniform plasmas can be controlled by the initial parameters of the laser pulse driver (the intensity, the spot size, the vacuum focal plane and the pulse duration) and the plasma density. However, it is challenging to derive an analytical expression to describe with good accuracy how the injection depends on these parameters. In instead we rely on simulations to show how the injection varies when different laser spot sizes are used. In Fig. \ref{fig: w0 scan},  we see that the plasma wake expands and injection occurs for three cases near the matched spot size, $k_\mathrm{p0}w_0$=3.5, 4 (matched) and 4.5, for fixed intensity ($a_0=4$) and thus different laser powers. When a laser pulse with a smaller spot size is used, the injection distance shrinks. The isolated injection that occurs at the end of the upramp around $k_\mathrm{p0}z=20$ for $k_\mathrm{p0}w_0=4$ and 4.5 is absent when $k_\mathrm{p0}w_0=3.5$. Note that the aforementioned injection of ultrahigh quality beams also occurs for laser pulse drivers with spot sizes far away from the matched spot size. We choose nearly matched laser spot sizes in this paper for the subsequent acceleration since a nearly matched laser pulse can be guided over many Rayleigh lengths of distance in plasma \cite{PhysRevSTAB.10.061301}.

The motion of plasma ions \cite{rosenzweig2005effects, an2017ion} which becomes important when the parameter $\Omega_\mathrm{b} \frac{\sigma_\mathrm{zb}}{c} = \sqrt{ \frac{n_\mathrm{b}}{n_\mathrm{p0}} \frac{m}{m_\mathrm{i}} } k_\mathrm{p0}\sigma_\mathrm{zb}$ exceeds unity would modify the distribution of the acceleration and focusing fields inside the wake and degrade the beam quality. Here $\Omega_\mathrm{b} = \frac{n_\mathrm{b} e^2}{m_\mathrm{i}\epsilon_0}$ is the ion plasma frequency for the beam density, $m_\mathrm{i}$ is the ion mass, $n_\mathrm{b}$ and $\sigma_\mathrm{zb}$ are the peak density and the duration of the injected beam. For the $g=0$ case, this value is $ \Omega_\mathrm{b} \frac{\sigma_\mathrm{zb}}{c} \sim 0.1$ even for the lightest hydrogen ions, so the ion motion is weak. Simulations performed with mobile hydrogen ions confirm that ion motion has little effect on the injected beam quality. 

We emphasize that fine grid sizes and advanced field solvers are necessary to model the ultrahigh quality electron beams generation in the highly nonlinear plasma wakes. Due to their ultra-low emittance and the ultra-strong focusing fields inside the ion channel, the injected beams are tightly focused down to spot sizes of $\sim0.1~k_\mathrm{p0}^{-1}$ with peak densities as high as $10^3\sim10^4~n_\mathrm{p0}$. Thus a fine grid size is needed to resolve them. When these high-density relativistic electrons propagate on the numerical grids, they can excite unphysical numerical fields since the grids can be viewed as a medium with a complicated dielectric tensor. Two important kinds of the unphysical effects are numerical Cherenkov radiation \cite{godfrey1974numerical, godfrey2013numerical, xu2013numerical} and numerical space-charge fields \cite{xu2020numerical}. These unphysical fields can modify the beams' evolution and degrade their qualities. Maxwell field solvers based on spectral solvers \cite{yu2014modeling, yu2015elimination, yu2015mitigation, yu2016enabling, godfrey2015improved} or with finite difference solvers with extended stencils \cite{xu2020numerical, li2021new} have recently been developed to suppress some or combinations of these numerical fields. In the aforementioned simulations, a combination of fine grid sizes ($\mathrm{d}z=\mathrm{d}r=\frac{1}{512}k_\mathrm{p0}^{-1}$) and the recently developed Xu solver \cite{xu2020numerical} are used to model the injected electrons with high-fidelity. For comparison, results with the Yee solver and/or coarse resolutions are presented in the Appendix A.

\section{III. Generation of ultrahigh brightness and ultrahigh current beams with hundreds of MeV}
\label{sec. ultrabright}
The properties of self-injected beams are determined by the plasma density and the evolution of the laser driver in the plasma in terms of both the $\xi$ and propagation distance variables. Thus, there is a large parameter space to explore and electron beams with different properties can be injected. In the previous section, a GeV-class high-quality beam with $10^{19}\sim10^{20}~\ampere/\meter^2/\rad^2$ brightness is produced by focusing a 4.3 J laser pulse into a plasma with $n_\mathrm{p0}=2\times 10^{18}~\centi\meter$. We show two more representative cases in this section: one is the generation of $\sim 100$ MeV beams with $\sim10^{21}~\ampere/\meter^2/\rad^2$ and the other is generation of beams with $\sim100~\kilo\ampere$ current. 

\begin{figure}[bp]
\includegraphics[width=0.5 \textwidth]{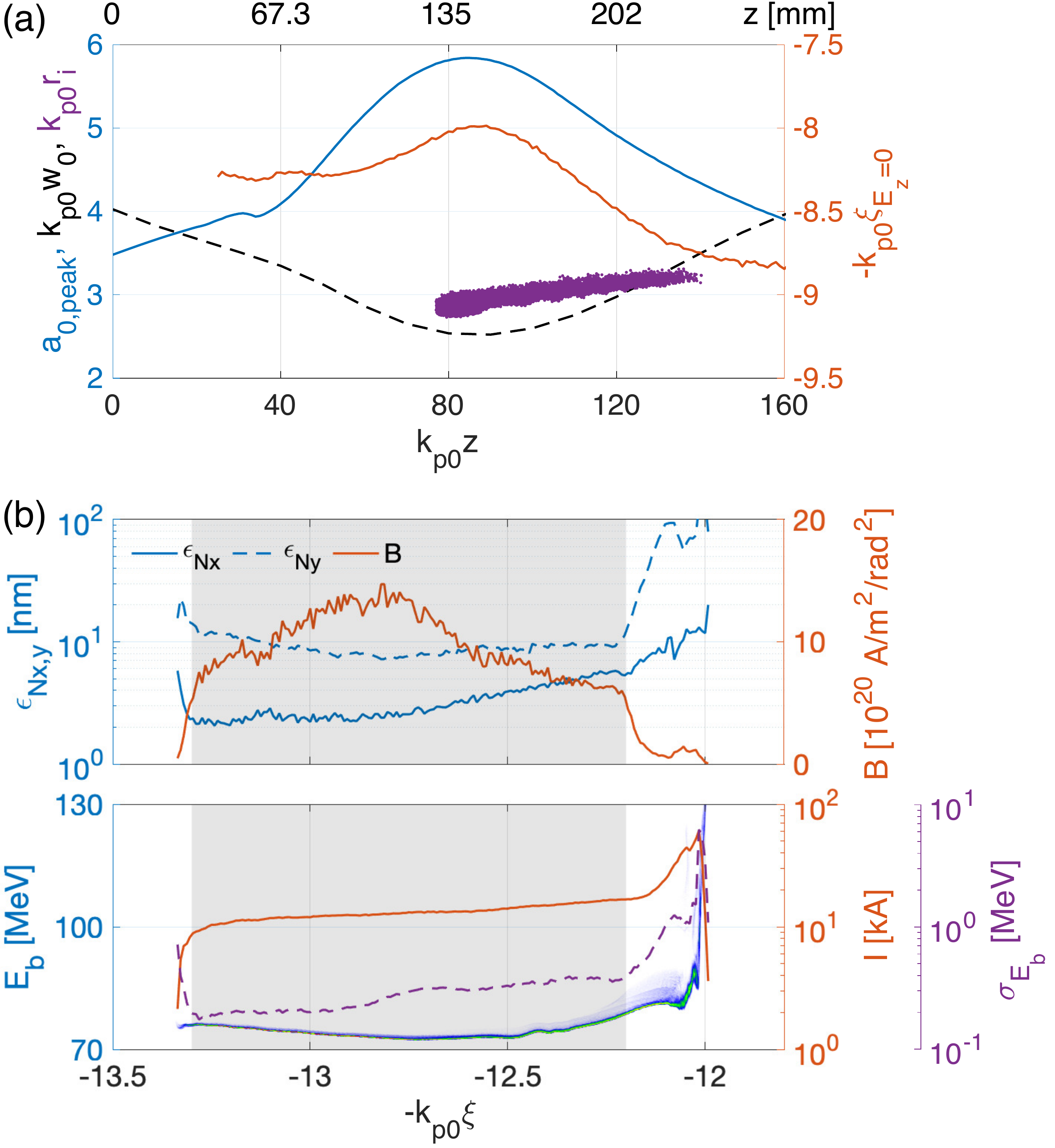}
\caption{\label{fig: ultra-bright} Generation of electron beams with $\sim10^{21}~\ampere/\meter^2/\rad^2$ brightness.  (a) The evolution of the normalized vector potential (blue), the projected spot size (black) of the laser diver, and the expansion of the wake (red). The purple dots show the initial axial and radial positions of the injected electrons. (b) The longitudinal phase space and the slice parameters of the injected beam at $k_\mathrm{p0}z=194$: the normalized emittance, the brightness, the current and the slice energy spread. The shadowed region represents the beam core.}
\end{figure}

Each simulation corresponds to a family of physical instances where the normalized parameters remain fixed. Thus, the emittance of the injected beams in downramp injection and self-injection in a uniform plasma scales with the background plasma density as $\epsilon_\mathrm{N} \propto k_\mathrm{p0}^{-1} \propto n_\mathrm{p0}^{-1/2}$ in beam driven plasma wakefield accelerators if the normalized dimensions of the beam drivers ($k_\mathrm{p0}\sigma_{x,y,z}$) and their normalized peak density ($\frac{n_\mathrm{b}}{n_\mathrm{p0}}$) are assumed fixed \cite{xu2017high}. This scaling for $\epsilon_\mathrm{N}$ also holds for laser-driven wakes if laser parameters are also scaled, including the laser frequency. However, as high power lasers are presently available in a limited range of wavelengths, then as the density changes the frequency ratio ($\frac{\omega_\mathrm{0}}{\omega_\mathrm{p0}}$) will not be scaled appropriately. At lower frequency ratios the laser evolves more rapidly and thus the scaling with $\epsilon_\mathrm{N}$ with density will be approximate. This scaling thus indicates that ultra-bright beams can be produced using a high density plasma for lasers as well as particle beam drivers. To confirm this, results from a simulation with $n_\mathrm{p0}=10^{19}~\centi\meter^{-3}$ are shown in Fig. \ref{fig: ultra-bright}. An 800 nm laser pulse with duration $\tau_\mathrm{FWHM}=25.9~\femto\second~(4.6\omega_\mathrm{p0}^{-1})$ and linear polarization in the $x$ direction is focused inside the plasma at [$z_\mathrm{f}=67.3~\micro\meter~(40k_\mathrm{p0}^{-1})$] with a focal spot size $w_0=5.9~\micro\meter~(3.5k_\mathrm{p0}^{-1})$. The laser contains 0.35 J energy and its peak power is $18.7~\tera\watt$. Its normalized vector potential at the vacuum focal plane is $a_0=4$. The spot size and the focal position \cite{PhysRevX.10.041015, PhysRevAccelBeams.25.101301} have a significant effect on the laser evolution and thus where the injection happens. These parameters are first scanned  using simulations with low resolution to find optimal operating parameters. 

As shown in Fig. \ref{fig: ultra-bright}(a), the laser is focused down even after its vacuum focal plane due to the self-focusing and starts to expand around $z\approx 135~\micro\meter~(80k_\mathrm{p0}^{-1})$. Injection starts at $z\approx 135~\micro\meter$ where the $a_0$ of the laser is focused to $a_0\approx 6$. The injection ceases at $z\approx 235~\micro\meter~(140k_\mathrm{p0}^{-1})$ due to a combination of beam loading from the injected electrons and the decrease of the laser intensity. Compared with the $g=0$ case in Fig. \ref{fig: concept}, the wake expands with a faster velocity of $0.01\sim 0.02 c$ which leads to a smaller energy chirp at the end of the injection [31.4 $\mega\electronvolt/(k_\mathrm{p0}\xi)$]. This small chirp is compensated quickly by the slope of the acceleration gradient thus the energy at the optimized acceleration distance is much lower than for the $g=0$ case in Fig. \ref{fig: concept}. 

Simulations show the core of the injected beam ($12.2<k_\mathrm{p0}\xi<13.3$ with $B\geq 0.5\times 10^{21}~\ampere/\meter^2/\rad^2$) achieves its minimum projected energy spread (1.6 MeV) at $z=326~\micro\meter~(194k_\mathrm{p0}^{-1})$ with an average energy 74.4 MeV. The slice energy spread is $\sim0.3$ MeV and the current is $\sim$13 kA. The charge contained in the beam core is 81 pC. As shown in Fig. \ref{fig: ultra-bright}(b), the emittance of the beam core is $\sim$3 nm ($x$), $\sim$9 nm ($y$) and its peak brightness reaches $1.5\times 10^{21}~\ampere/\meter^2/\rad^2$. The unequal emittance (and spot sizes) along the two transverse directions of the injected beam can be traced back to the fact that only electrons with initial azimuthal angles around $\theta_i \equiv \mathrm{atan2}(y_i, x_i)=\pm \frac{\pi}{2}$ are injected and the details are presented in Appendix C. The dephasing between the injected electrons and the laser driver sets a limit on the maximum energy gain of the injected beam as $\sim100$ MeV in a plasma with $\sim10^{19}~\centi\meter^{-3}$ density \cite{PhysRevSTAB.10.061301}. Injection and acceleration in such a high-density plasma can serve as an injector with 100 MeV-class ultra-bright electron beams. 

\begin{figure}[bp]
\includegraphics[width=0.5 \textwidth]{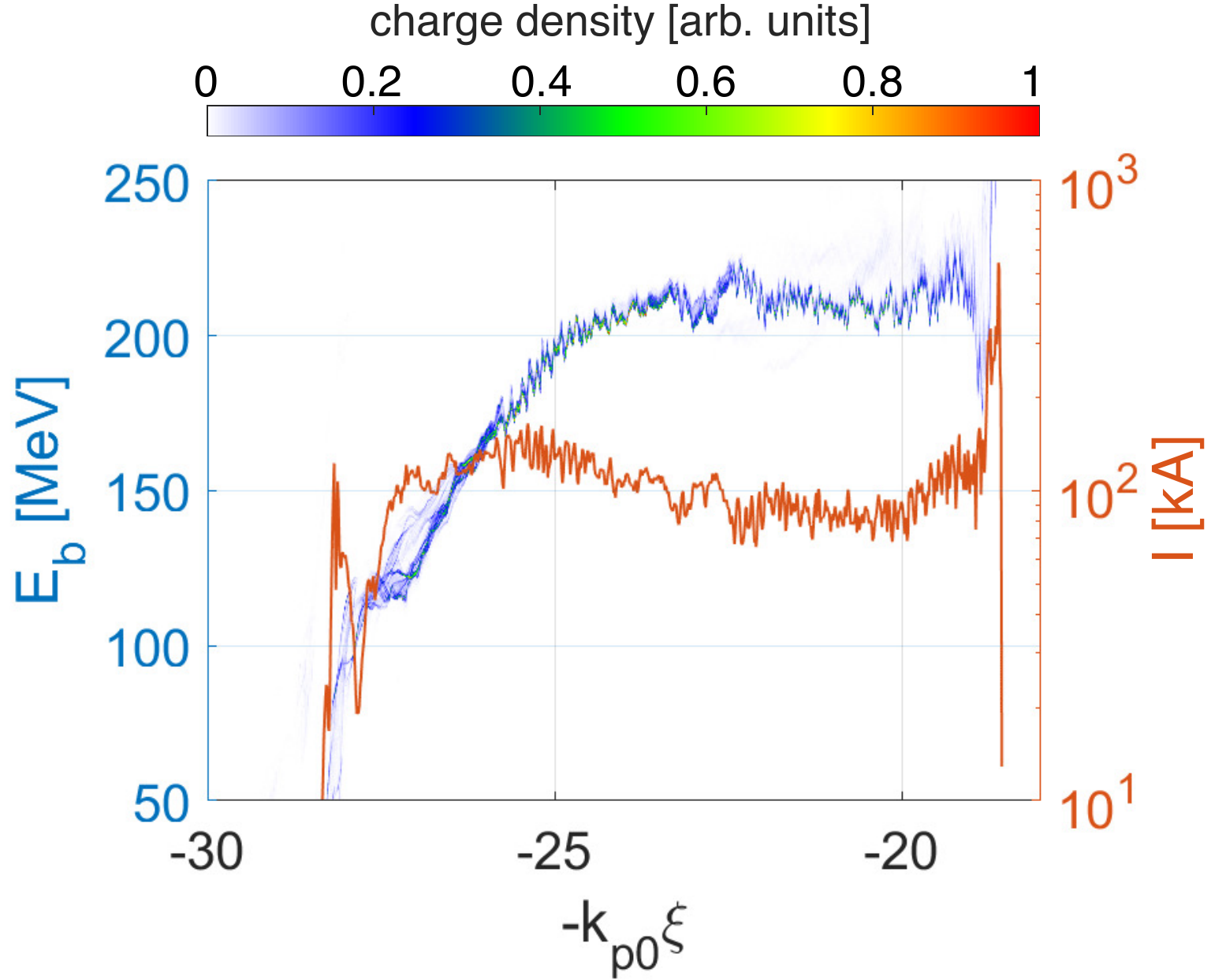}
\caption{\label{fig: 100 kA} Generation of electron beams with $\sim100~\kilo\ampere$ current by sending a $1.2~\peta\watt$ laser pulse into a plasma with $10^{19}~\centi\meter^{-3}$ density: the longitudinal phase space and the current profile at $\omega_\mathrm{p0}t = 400$.}
\end{figure}

Based on the empirical relation between the current of the injected beams and the $a_0$ of the laser pulse drivers, beams with hundreds of kA are expected to be produced when using laser pulses with $a_0\gtrsim 10$. We present simulation results in Fig. \ref{fig: 100 kA} for a case where a PW laser pulse and plasma density $n_\mathrm{p0}=10^{19}~\centi\meter^{-3}$ are used. An 800 nm, linearly polarized ($x$) laser pulse with duration $\tau_\mathrm{FWHM}=38.8~\femto\second~(6.9\omega_\mathrm{p0}^{-1})$ is focused outside of the plasma [$z_\mathrm{f}=-640~\micro\meter~(-380k_\mathrm{p0}^{-1})$] with a focused spot size $w_0=11.8~\micro\meter~(7k_\mathrm{p0}^{-1})$. The laser contains 33.4 J energy and its peak power is $1.2~\peta\watt$. Its normalized vector potential at its focal plane is $a_0=16$. 

Self-injection starts at $\sim100 k_{\mathrm{p0}}^{-1}$ and ends at $\sim300 k_{\mathrm{p0}}^{-1}$. A beam with $\sim60~\femto\second$ duration and $\sim$100 kA current is formed at $\omega_\mathrm{p0}t=400$. The total injected charge is $\sim6~\nano\coulomb$ which can be further improved if a plasma downramp is introduced to elongate the injection distance. Beams with 10s of MA current may be obtained by controlling the acceleration distance to form an energy chirp along this beam and compress it in a small chicane \cite{emma2021terawatt}. A high-fidelity simulation of the generation of this high current beam requires a resolution finer than $\frac{1}{512}k_\mathrm{p0}^{-1}$. However, a resolution with $\frac{1}{256}k_\mathrm{p0}^{-1}$ is used in Fig. \ref{fig: 100 kA} due to the significant computational cost to model the large plasma wave wake, and this leads to a highly suspect emittance ($\sim\micro\meter$) and artificial ripples in the longitudinal phase space of the beam (see Fig. \ref{fig: 100 kA}). Based on our experience for other cases where we are able to carry fully resolved simulations, we believe that these effects are numerical and not physical.

\section{IV. Discussion on physical effects which would degrade the beam quality}
The aforementioned results have demonstrated the ability to inject ultrahigh quality electron beams in the simplest configuration of laser wakefield accelerators. However, the beam quality reported in experiments (either based on a density downramp or the evolution of the laser pulse driver) are significantly worse than the predictions of PIC simulations. In this section we discuss the possible reasons behind this large gap.

The transverse deceleration process which results in injection of low-emittance beams depends on having a well defined sheath and the axial symmetry of the nonlinear wake. Thus, any factors that affect the sheath properties and the symmetry of the wake degrade the emittance and brightness of the injected beams. For instance: non diffraction-limited lasers, a finite temperature of the plasma electrons \cite{xu2017high, PhysRevAccelBeams.22.111301}, an asymmetric driver (intensity and/or phase distortions) \cite{Glinec_2008, Mangles2009APL, PhysRevLett.105.215001, PhysRevX.5.031012}, a finite transverse bulk velocity of a gas produced from a gas jet, transversely non-uniform plasma density distribution at spatial scales smaller or larger than plasma skin depth \cite{Stupakov2017undulator, Ma2018Angular}. These imperfections not only affect the injection process but also degrade the beam quality during the subsequent acceleration by modifying the linear focusing field and the transversely uniform acceleration field in an axisymmetrical nonlinear wake. However, it is not straightforward to study these effects in the Q3D geometry since significant effort is needed to understand how many modes are necessary to model these asymmetry wakes with high-fidelity. Meanwhile, the computational costs of full 3D simulations are currently unaffordable even using GPU based hardware. Therefore, we will focus on the effects of the plasma temperature in this section. 

\begin{figure}[bp]
\includegraphics[width=0.5\textwidth]{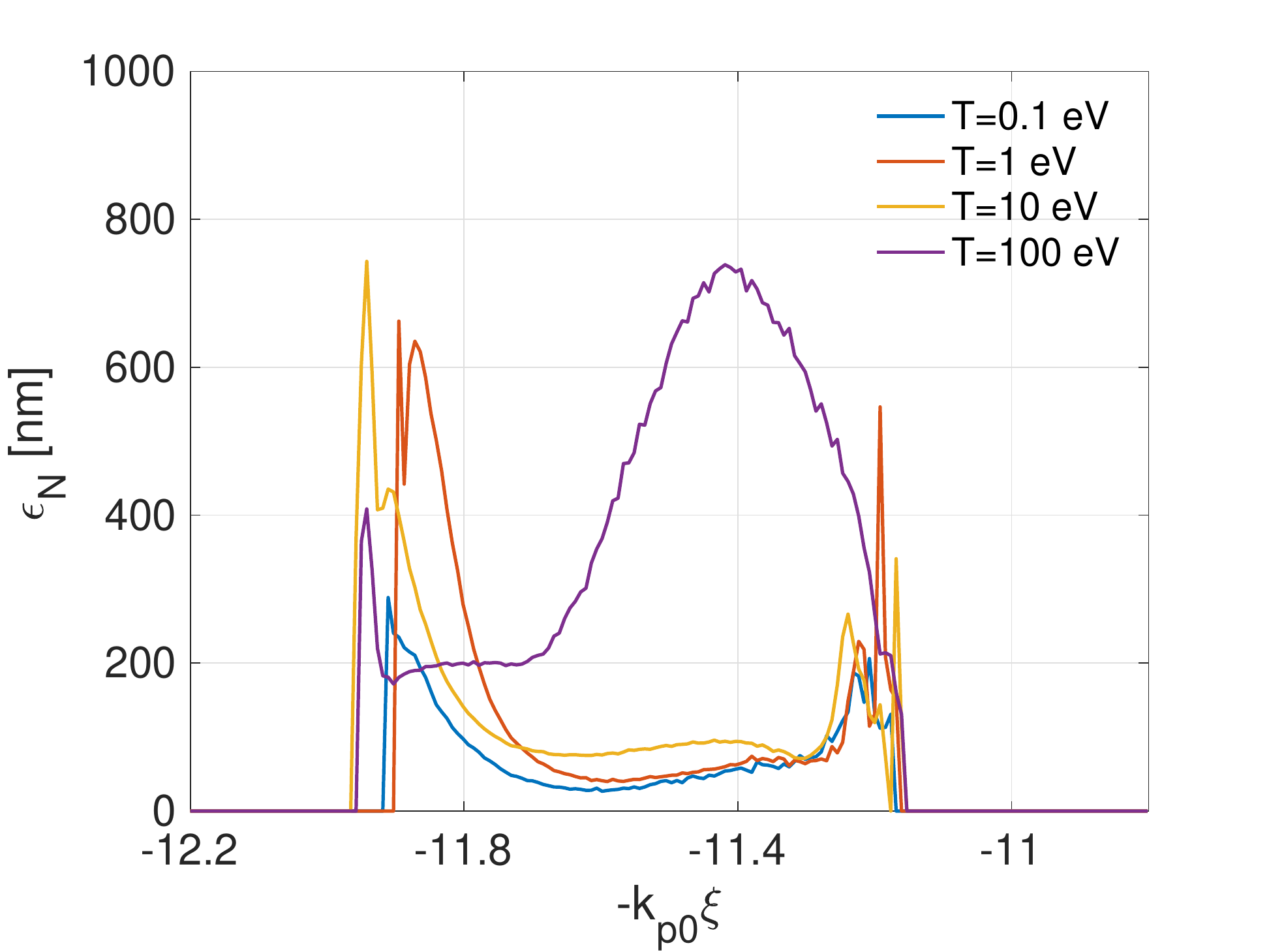}
\caption{\label{fig: T scan} Comparison of the emittance of the self-injected electron beams under different plasma electron temperature. The parameters of the laser pulse and the plasma are the same as that of Fig. \ref{fig: concept}.}
\end{figure}

In practical experiments, the plasma is created in many ways, e.g., high-voltage discharge or optical ionization by a low energy laser pulse. Depending on the ionization process, the plasma electrons can will be distributed with different temperatures which may affect the quality of the injected electron beams. In the self-injection results shown in Sec. II, we assumed an initial plasma electron temperature of $T=0.1~\electronvolt$. In Fig. \ref{fig: T scan}, we compare the emittance of the self-injected beams from plasmas with higher temperatures ($1~\electronvolt$, $10~\electronvolt$ and $100~\electronvolt$). While the emittance within the beam cores are similar for $T=1~\electronvolt$ and $T=0.1~\electronvolt$, it grows by a factor of $\sim$2 for $T=10~\electronvolt$. The emittance increases dramatically and approaches $\sim\micro\meter$ when the plasma electron temperature is $T=100~\electronvolt$. This indicates that eV-level plasma temperatures are necessary for generation of high-quality beams with nanometer-scale emittances.

Even when these ultra-bright beams are produced in the plasma, the subsequent transport may also degrade the emittance due to the mismatch of the beams' transverse phase space \cite{PhysRevSTAB.15.111303, Antici2012JAP, PhysRevSTAB.16.011302, PhysRevSTAB.17.054402, PhysRevSTAB.18.041302,  xu2016matching, PhysRevAccelBeams.23.011302, PhysRevAccelBeams.22.041304}, especially when other macro parameters of the beams (e.g., energy, pointing angles and transverse positions) fluctuate from shot-to-shot \cite{labat2018robustness}. Thus, carefully designed transport stages are additionally necessary for utilization of high-quality self-injected beams from PBA.

\section{V. Conclusions}
We have shown that the dynamics of the self-injection in uniform plasmas is the same as that of density downramp injection \cite{xu2017high} and its variants \cite{PhysRevAccelBeams.23.021304, li2022ultra}, thus beams with $10^{20}\sim 10^{21}~\ampere/\meter^2/\radian^2$ and $0.1\sim 1$ MeV slice energy spreads can be produced. Fine grid sizes and advanced field solvers are necessary to model this ultrahigh quality beams generation in PIC codes. 

Due to its low emittance, the GeV beam described in Sec. II is focused tightly to a spot size as small as $240~\nano\meter$ and reaches a peak density as high as $2\times 10^{21}~\centi\meter^{-3}$. The generated beams with higher brightness or current in Sec. III may be focused to an even higher beam density if their energy is boosted to GeV-levels. These extremely dense beams can enable many novel applications, such as driving all-optical XFELs \cite{xu2023ultra}, generating ultrabright $\gamma$-rays through a beam-plasma instability \cite{benedetti2018giant} or beam-multifoil collisions \cite{PhysRevLett.126.064801}, driving plasma wakefield accelerators with $\sim10$ TeV/m acceleration gradient in a solid density plasma \cite{sahai2020nanostructured} and studying strong-field QED with beam-beam collisions \cite{PhysRevLett.122.190404} or beam-plasma collisions \cite{matheron2022probing}. 

Using high-fidelity large-scale numerical simulations, we have shown that beams with extreme parameters can be produced when a laser pulse propagates and evolves in a uniform plasma. This approach can dramatically simplify the complexity of plasma-based accelerators. Our findings may also stimulate future research to study why current experiments can not deliver these ultrahigh quality beams and on what is needed to finally produce these beams in experiments to enable many novel plasma acceleration driven applications.

\section{acknowledgments}
This work was supported by the Fundamental Research Funds for the Central Universities, Peking University, the National Natural Science Foundation of China (NSFC) (No. 11921006) and the National Grand Instrument Project (No. 2019YFF01014400), the U.S. Department of Energy under Contracts No. DE-SC0010064, the U.S. National Science Foundation under Grants No. 2108970, and the DOE Scientific Discovery through Advanced Computing (SciDAC) program through a Fermi National Accelerator Laboratory (FNAL) subcontract No. 644405. The simulations were performed on the resources of the National Energy Research Scientific Computing Center (NERSC), a U.S. Department of Energy Office of Science User Facility located at Lawrence Berkeley National Laboratory.

\appendix
\section{Appendix A: Comparison of the injected beams from simulations with different resolutions and solvers}
\label{sec. comparison}

\begin{figure*}
\includegraphics[width=1.0\textwidth]{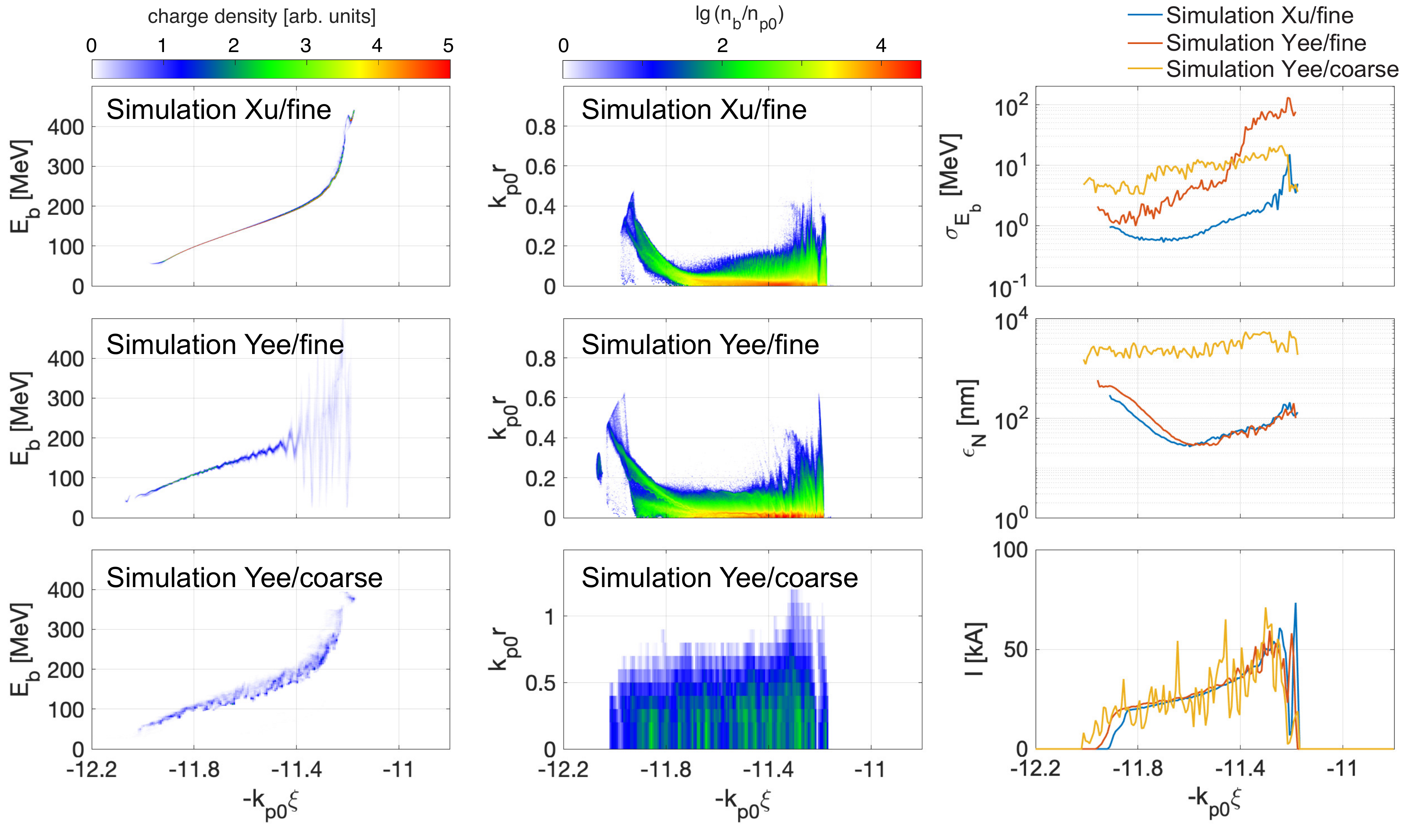}
\caption{\label{fig: 3 simulations} Comparison of the injected beams at $\omega_\mathrm{p0}t=250$ in a uniform plasma from simulations with different resolutions and different solvers. Simulation Xu/fine: $\mathrm{d}z=\mathrm{d}r=\frac{1}{512}k_\mathrm{p0}^{-1}$ and the Xu solver; Simulation Yee/fine: $\mathrm{d}z=\mathrm{d}r=\frac{1}{512}k_\mathrm{p0}^{-1}$ and the  Yee solver; Simulation Yee/coarse: $\mathrm{d}z=\frac{1}{150}k_\mathrm{p0}^{-1}, \mathrm{d}r=\frac{1}{10}k_\mathrm{p0}^{-1}$ and the Yee solver.}
\end{figure*}

In Sec. II, we showed that the beam injected in a uniform plasma shares the same dynamics as in a plasma downramp and is characterized by ultrahigh brightness and low slice energy spread. However, the beam quality from simulations is closely related to the simulation setup, such as the grid sizes and the Maxwell field solver. In Sec. II, a combination of fine grid sizes ($\mathrm{d}z=\mathrm{d}r=\frac{1}{512}k_\mathrm{p0}^{-1}$) and the recently developed Xu solver \cite{xu2020numerical} are used to model the injected electrons with high-fidelity, which will be referred as `Simulation Xu/fine' in the following discussions. In Fig. \ref{fig: 3 simulations}, we show that the qualities of the injected beams in simulations with coarse resolutions or the Yee solver are much worse than the results from Simulation Xu/fine.

\begin{figure*}
\includegraphics[width=1.0\textwidth]{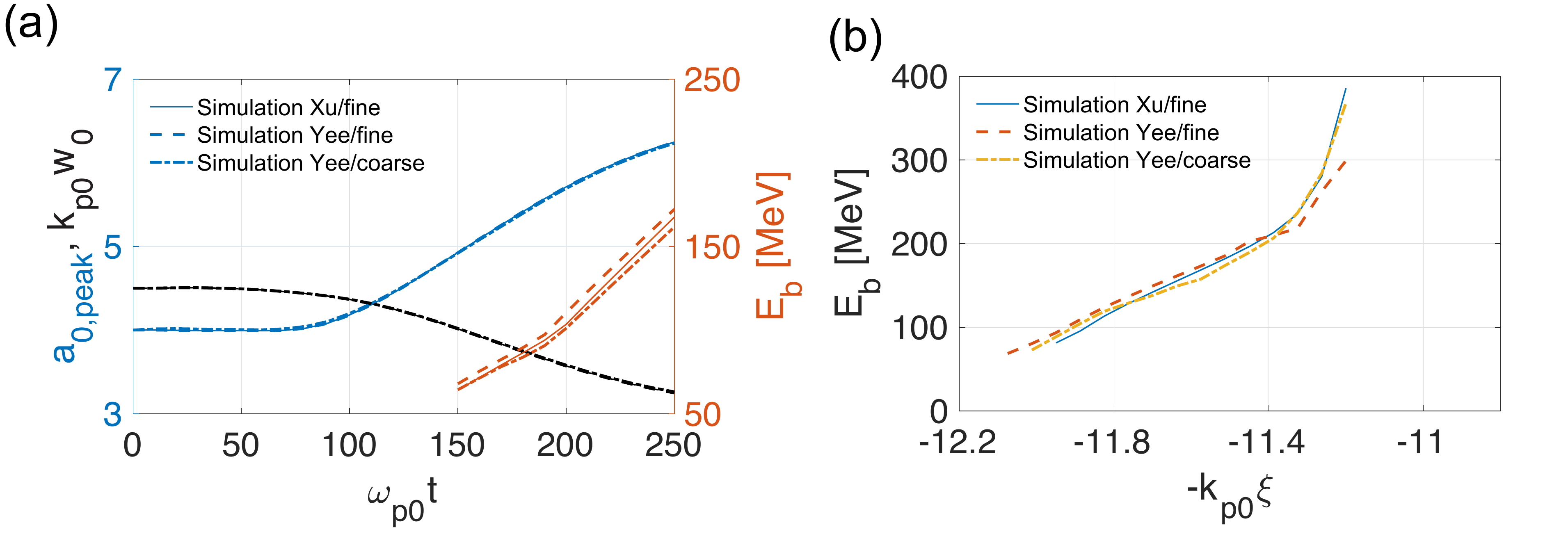}
\caption{\label{fig: 3 simulations macro} (a) Comparison of the evolution of the laser, the average energy of the injected electrons from simulations with different settings. (b) The energy of the electrons at different slices ($\frac{1}{16}k_\mathrm{p0}^{-1}$ width) from these simulations at $\omega_\mathrm{p0}t=250$. }
\end{figure*}

Simulation Yee/fine uses the same setup of Simulation Xu/fine except the Yee field solver \cite{yee1966numerical} is used instead of the Xu solver. Simulation Yee/coarse uses a much coarser grid resolution ($\mathrm{d}z=\frac{1}{150}k_\mathrm{p0}^{-1}, \mathrm{d}r=\frac{1}{10}k_\mathrm{p0}^{-1}$) and the Yee solver with 128 macro-particles per cell to improve the statistics. The time step of $\mathrm{d}t = 0.00625 \omega_\mathrm{p0}^{-1}$ is close the Courant limit. The longitudinal phase space and the real space distribution of the injected beams in these three simulations are compared in Fig. \ref{fig: 3 simulations}. In Simulation Yee/fine, the beam is focused to a similar real space distribution as in Simulation Xu/fine. However, these high-density electrons excite unphysical high-frequency electromagnetic fields with large amplitude \cite{xu2020numerical} when the Yee solver is used. These unphysical fields then modulate the longitudinal phase space of the beam and lead a much larger slice energy spread (several $\sim 100~\mega\electronvolt$). The beam in Simulation Yee/coarse has a large emittance ($>1~\micro\meter$) and thus a large spot size. Since the electrons are not tightly focused, their density is much lower than in Simulation Xu/fine and Yee/fine and they excite numerical fields with lower amplitude than in Simulation Yee/fine. Thus the energy modulation is not as severe as in Simulation Yee/fine and the slice energy spread is several $\sim 10~\mega\electronvolt$. The total charge of the injected beams in all three simulation is very similar but the current profiles in Simulations Yee/fine and Yee/coarse are modulated as a result of the slippage between the electrons with unphysical energy modulation. 

As shown in Fig. \ref{fig: 3 simulations}, the simulation resolution and field solver can significantly impact the quality of the injected beams due to the subtle self-interactions between the high density particles. However, we want to stress that quantities that depend on physics on larger spatial scales are not sensitive to the choices of the grid sizes and the solver, such as the evolution of the laser driver and the average energy of the injected beam as shown in Fig. \ref{fig: 3 simulations macro}(a). We also compare the slice mean energy of the injected beams in these simulations in Fig. \ref{fig: 3 simulations macro}(b). When the duration of the slice is chosen as $\frac{1}{16}k_\mathrm{p0}^{-1}$ which is longer than the wavelength ($\sim 0.03 k_\mathrm{p0}^{-1}$) of the unphysical energy modulation, the slice mean energy has similar profiles in the three simulations. This indicates simulations with low resolutions and the Yee solver can be used to predict the coarse energy space of the injected beam, and they may be recommended in some situations due to the much lower computational cost. Machine learning algorithms may be combined with these low cost simulations to predict these properties for injected electrons.

\appendix
\section{Appendix B: Simulation setup}
\label{sec:setup}
For the simulations shown in Sec. II, we use a moving window propagating at speed of light in vacuum $c$ with a box size of $14 k_\mathrm{p0}^{-1}\times 12 k_\mathrm{p0}^{-1}$ and $7168\times 6144$ cells along the $z$ and $r$ directions, respectively. The grid sizes are chosen as $\mathrm{d}z=\mathrm{d}r=\frac{1}{512}k_\mathrm{p0}^{-1} \approx 7.35~\nano\meter$ to resolve the tightly focused injected electrons while the time step is $\mathrm{d}t=\frac{1}{1024}\omega_\mathrm{p0}^{-1}\approx 12.2~\atto\second$ to satisfy the CFL condition. To represent the plasma electrons, we use 8 macro-particles per cell (ppc) where they are distributed at one $r-z$ location and 8 different values of the azimuthal angle $\theta$. All physical quantities in Q3D code \cite{lifschitz2009particle, davidson2015implementation} are decomposed into azimuthal modes $\mathrm{exp}(i m \theta)$, where $\theta$ is the azimuthal angle. Two modes with $m=0$ and $1$ are included in the simulations shown here to describe the linearly polarized laser pulses. The Xu solver \cite{xu2020numerical} is used to eliminate the numerical Cherenkov radiation \cite{godfrey1974numerical, godfrey2013numerical, xu2013numerical} and the numerical space charge field \cite{xu2020numerical} from the high-density relativistic injected electrons and model the injection and acceleration with high-fidelity. The laser's electric field has a symmetric temporal profile of $10\tau^3 - 15\tau^4 + 6\tau^5$, where $\tau=\frac{ \sqrt{2} (t-t_0) }{ \tau_\mathrm{FWHM}  }$ while the radius of the plasma column is $41.4~\micro\meter~(11k_\mathrm{p0}^{-1})$.

The simulation in Fig. \ref{fig: ultra-bright} uses a moving window with a box size of $16\times 16~(c/\omega_\mathrm{p0})^2$ whose grid sizes are $\mathrm{d}z=\mathrm{d}r=\frac{1}{512}k_\mathrm{p0}^{-1} \approx 3.29~\nano\meter$. The time step is $\mathrm{d}t=\frac{1}{1024}\omega_\mathrm{p0}^{-1}\approx 5.48~\atto\second$ and 8 macro-particles per cell (distributed at 8 different values of $\theta$ at one $r-z$ location) are used to represent the plasma electrons. The simulation in Fig. \ref{fig: 100 kA} uses a larger moving window with a box size of $32k_\mathrm{p0}^{-1}\times 32k_\mathrm{p0}^{-1}$ to model the larger plasma wave wake excited by the more intense laser pulse. Limited by computational cost, a resolution with $\mathrm{d}z=\mathrm{d}r=\frac{1}{256}k_\mathrm{p0}^{-1} \approx 6.58~\nano\meter$ and a time step $\mathrm{d}t=\frac{1}{512}\omega_\mathrm{p0}^{-1}\approx 10.96~\atto\second$ is used. We still use two modes with $m=0$ and 1 in the simulations and the Xu solver. We use 8 macro-particles per cell (distributed at 8 different values of $\theta$ at one $r-z$ location) for the plasma electrons. The plasma electrons in both simulations have an initial $0.1~\electronvolt$ temperature. Note $n_\mathrm{p0}=2\times 10^{18}~\centi\meter^{-3}$ in Sec. II and $n_\mathrm{p0}=10^{19}~\centi\meter^{-3}$ in Sec. III.

\appendix
\section{Appendix C: Initial azimuthal distribution of the injected electrons}

\begin{figure*}
\includegraphics[width=1.0\textwidth]{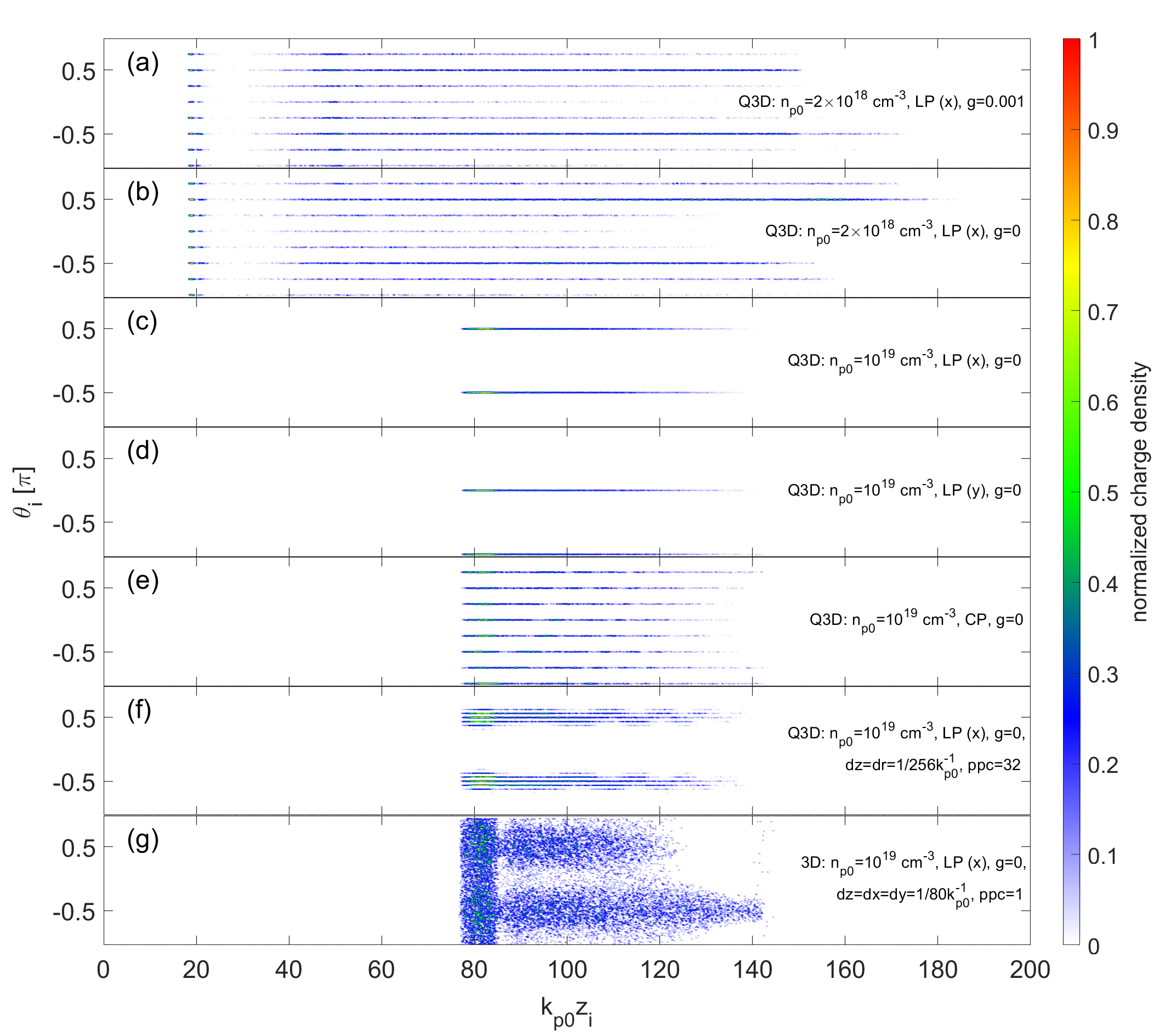}
\caption{\label{fig: angles} The charge density distribution of the injected electrons in the $\theta_i-z_i$ plane. (a) Q3D simulation for $n_\mathrm{p0}=2\times 10^{18}~\centi\meter^{-3}$ and $g=0.001$. (b) Q3D simulation for $n_\mathrm{p0}=2\times 10^{18}~\centi\meter^{-3}$ and $g=0$. (c) Q3D simulation for $n_\mathrm{p0}= 10^{19}~\centi\meter^{-3}$ and $g=0$. (d) Q3D simulation for $n_\mathrm{p0}= 10^{19}~\centi\meter^{-3}$, $g=0$ and the laser pulse driver is linearly polarized along $y$-direction. (e) Q3D simulation for $n_\mathrm{p0}= 10^{19}~\centi\meter^{-3}$, $g=0$ and the laser pulse driver is circularly polarized. (f) Q3D simulation with low resolution and more macro-particles per cell for $n_\mathrm{p0}= 10^{19}~\centi\meter^{-3}$ and $g=0$. (g) 3D simulation for $n_\mathrm{p0}= 10^{19}~\centi\meter^{-3}$ and $g=0$. Note the peak intensity of the circularly polarized laser pulse is reduced by a factor of $\sqrt{2}$ to ensure it has the same ponderomotive force as the linearly polarized laser.}
\end{figure*}

When the plasma density is much lower than the critical density of the laser driver and the laser pulse consists of many cycles, the ponderomotive approximation is valid and the plasma wake is nearly axisymmetric. As a result, the initial positions of the injected electrons are distributed uniformly along the azimuthal direction. As the plasma density increases, a shorter laser pulse driver is needed to excite the wake effectively, thus the oscillatory laser electric field starts to play a role and the injection becomes non-uniform azimuthally which leads to a transversely asymmetric beam, i.e., the beam has different spot sizes and emittance along the two transverse directions. 

The initial azimuthal angle $\theta_i$ of the injected electrons for different cases are shown in Fig. \ref{fig: angles}. When $n_\mathrm{p0}=2\times 10^{18}~\centi\meter^{-3}$, there are electrons injected from all angles for both the  $g=0$ and $g=0.001$ cases. Although their distribution with the initial angle is not uniform, i.e., more electrons originate from $\theta_i=\pm \frac{\pi}{2}$ [Fig. \ref{fig: angles}(a)-(b)],  the distribution is approximately symmetric between $k_\mathrm{p0}z_i\approx 40$ and $\sim140$. At the end of the injection ($k_\mathrm{p0}z_\mathrm{i} \gtrsim 140$), the angular distribution becomes asymmetric which leads to off-axis beam tails as shown in Fig. \ref{fig: downramp and no ramp}. The reason behind this asymmetric angular distribution has not been understood thoroughly. Note that 8 macro-particles are initialized azimuthally with angles $\left(-\pi, -\frac{3}{4}\pi, -\frac{1}{2}\pi, -\frac{1}{4}\pi, 0, \frac{1}{4}\pi, \frac{1}{2}\pi, \frac{3}{4}\pi \right) $ in these Q3D simulations.

When $n_\mathrm{p0}=10^{19}~\centi\meter^{-3}$ and a laser pulse driver with fewer cycles is used, only electrons originating from two angles, $\theta_i=\pm \frac{\pi}{2}$, are injected as shown \ref{fig: angles}(c). Since the angular distribution is still symmetric about the $x$ and $y$ axes, the injected beam is characterized by a low emittance. The values of the injected angles depends on the polarization direction of the laser driver as shown in Fig. \ref{fig: angles}(c)-(d). When a circularly polarized laser is used, the augular distribution becomes much more uniform. This concentration on initial angles  when a linearly polarized laser pulse is used is confirmed in a Q3D simulation with lower resolution ($\mathrm{d}z=\mathrm{d}r=\frac{1}{256}k_\mathrm{p0}^{-1}$) but more macro-particles per cell (f) and a full-3D simulation (g) with lower resolution ($\mathrm{d}z=\mathrm{d}x=\mathrm{d}y=\frac{1}{80}k_\mathrm{p0}^{-1}$).

\bibliography{refs_xinlu}

\end{document}